\documentclass{ws-ijmpa}
\usepackage[super,compress]{cite}
\usepackage{graphicx}
\usepackage{color}\usepackage{graphicx,graphics}\usepackage{multirow}\usepackage{float}\usepackage{ulem}
\usepackage[pdfstartview=FitH]{hyperref}
\usepackage{setspace}\usepackage{slashed}\usepackage{booktabs}
\usepackage{mathrsfs} \usepackage{braket}
\hypersetup{colorlinks=true, citecolor=blue, linkcolor=red, filecolor=black,urlcolor=blue}

\begin{document}
\markboth{Kai Zhu}{Triangle relations for XYZ states}

%
\catchline{}{}{}{}{}
%

\title{Triangle relations for XYZ states}

\author{Kai Zhu}

\address{Institute of High Energy Physics, Beijing 100049, China\footnote{
zhuk@ihep.ac.cn}}

\maketitle

\begin{history}
\received{Day Month Year}
\revised{Day Month Year}
\end{history}

\begin{abstract}
We propose novel triangle relations, not the well-known triangle singularity, for better understanding of the exotic XYZ states. Nine XYZ resonances,  $X(3872)$, $Y(4230)$, $Z_c(3900)$, $X(4012)$, $Y(4360/4390)$, $Z_c(4020)$, $X(4274)$, $Y(4660)$, and $Z_{cs}(3985)$ have been classified into triples to construct three triangles based on the assumption that they are all tetra-quark states. Here $X(4012)$ has not been observed experimentally yet, and we predict its mass and roughly the width. We also suggest some channels deserving search for with high priority based on this hypothesis, as well as predictions of a few production/decay rates of these channels. We hope further experimental studies of the XYZ states will benefit from our results.

\end{abstract}

\ccode{PACS numbers:}


Charmonium-like sates, the so called XYZ particles, have been hot topics in high energy physics community since $X(3872)$ was discovered in 2003 by Belle~\cite{hep-ex/0309032}. Their special characters make them good candidates of multiple-quark states~\cite{1907.07583} beyond the conventional mesons (quark anti-quark pair) and baryons (triple quarks). In addition to be a novel material form, they also provide idea labs for studying the non-perturbative QCD, the theory describing strong interaction. Studying their production and decay rates individually, and comparing them to that of the charmonium states, should shed light on the internal structures and interaction mechanism of these exotic states. Then it will lead to better understanding of the strong interaction, such as color confinement, at the several $\mathrm{GeV}$ energy region. Various theoretical models have been proposed to interpret the experimental results and describe the nature of these XYZ states, including models, for example, tetra-quark, hybrid, conventional charmonium, molecular, and even cusp effect. For more details, our readers may read the fourth part of~\cite{1907.07583} and the references quoted in this review. However, in short, a global and satisfied frame, who can describe the whole XYZ states well, is still in absent. The situation of our knowledge on the multiple-quark states is somewhat similar to that on the glueballs, both of them are allowed by QCD 
but without experimentally conclusive evidence yet. Also, the mixing between these exotic states with normal mesons or baryons make the situation much more complex.

Recently, three results reported by BESIII collaboration have called our attention. Two of them are the confirmation of the radiative transition from $Y(4230)$ to $X(3872)$~\cite{1903.04695} and pion transition from $Y(4230)$ to $Z_c(3900)$~\cite{2004.13788}. The third one is the first observation of $Z_{cs}(3985)$~\cite{2011.07855}, it is based on the data samples of $e^+ e^-$ collision at the center-mass-energy close to $Y(4660)$. Even the statistic is low, it seems $Z_{cs}(3985)$ is produced by a kaon transition from $Y(4660)$. These new information and previous experimental results inspire us to investigate the relations between these XYZ states and suggest a general frame containing them. In this frame, XYZ states are classified into various sets, and each contains one X, one Y, and one Z states with similar quark components. So a single set can be illuminated by a triangle. Each point of a triangle is a state, while each side of the triangle is a hadronic or radiative transition. In this frame, nine XYZ states, $X(3872)$, $Y(4230)$, $Z_c(3900)$, $X(4012)$, $Y(4360/4390)$, $Z_c(4020)$, $X(4274)$, $Y(4660)$, and $Z_{cs}(3985)$ have been classified into three sets to construct triple triangles. We believe such a frame would provide insights on them, and can be a guide to future experimental searches and measurements.

The construction of a triangle start with a simple hypothesis, that the XYZ states are tetra-quark states and can be described universally in a similar quark components, {\it i.e.} $\ket{h}\ket{c\bar{c}}$. Here the $\ket{h}$ is a superimposition of light quark components, and same for $X$ and $Y$ states as
 \begin{equation}
  \ket{h}_{X/Y} = p_{h1} \ket{u\bar{u}} + p_{h2} \ket{d\bar{d}} + p_{h3} \ket{s\bar{s}} \;\;,
  \label{eq:hxy}
 \end{equation}
 where $p_{h1}$, $p_{h2}$, and $p_{h3}$ are amplitude strength can be determined as parameters via experimental measurements. However, the form of $\ket{h}$ for $Z$ states should be slightly different since their iso-spin is not zero, and can be written as
 \begin{equation}
 \label{eq:hz}
 \ket{h}_{Z} = p'_{h1} \ket{u\bar{d}} + p'_{h2} \ket{u\bar{s}} + p'_{h3} \ket{d\bar{u}} + p'_{h4} \ket{s\bar{u}} + p'_{h5} \ket{d\bar{s}} + p'_{h6} \ket{s\bar{d}} + p'_{h7} \ket{u\bar{u}} - p'_{h7} \ket{d\bar{d}} \;\; ,
\end{equation}
 where parameters $p'_h$s will take corresponding values for various states.
 Similarly $\ket{c\bar{c}}$  can be expanded into a superimposition of hidden-charm states as
 \begin{equation}
 \label{eq:cc}
 \ket{c\bar{c}} = p_{c1} \ket{\eta_c} + p_{c2} \ket{J/\psi} + p_{c3} \ket{\chi_{c0}} + p_{c4} \ket{\chi_{c1}} + p_{c5} \ket{\chi_{c2}} + p_{c6} \ket{h_c} + p_{c7} \ket{\eta'_c} + p_{c8} \ket{\psi'} + \cdots \;\;,
 \end{equation}
where only the chamonia below the open-charm threshold have been explicitly addressed, since we assume the contribution from the hidden-charm above the threshold is negligible in the cases we discussed. 
Notice our hypothesis is significantly different to the models with these XYZ states are pure charmonia or fake resonances from cusp effect.
From this hypothesis, a naive inference is that a mass of any XYZ state is mainly determined by the heavy quark component, and strongly affected by their couplings to open-charm channels. This inference would be used to determine the mass weights of these triangles. Since there are many different open-charm combinations, here we only consider the thresholds of the following S-wave open-charm channels, {\it i.e.} $DD_1(2420)$, $D^*D_1(2420)$, $D_s D_{s1}(2460)$, $D^*_s D_{s1}(2460)$, $D^*_s D_{s1}(2536)$, $D^* D_2(2460)$, and $D^*_s D_{s2}(2573)$. They are believed to provide the largest contributions. From a simple sum of the masses of the D mesons, we know the threshold are roughly at $4.29$ GeV, $4.43$ GeV, $4.43$ GeV, $4.57$ GeV, $4.65$ GeV, $4.47$ GeV, and $4.68$ GeV, respectively. It seems if there are seven triangles, two of them are according to the same $4.43$ GeV would be considered degeneracy. However, after considering the quark component, it is more likely the two pais $(D^* D_1 \mathrm{,} D^* D_2)$ and $(D^*_s D_{s1} \mathrm{,} D^*_s D_{s2})$ should be consider as degeneracies. So the reduced four thresholds are $4.29$ GeV, $4.43(4.47)$ GeV, $4.57$ GeV, $4.65(4.68)$ GeV. And the shifts between the excited ones to the fundamental one would be $140$ MeV, $280$ MeV, $360$ MeV.
\begin{figure}[htb]
\includegraphics[width=0.95\textwidth]{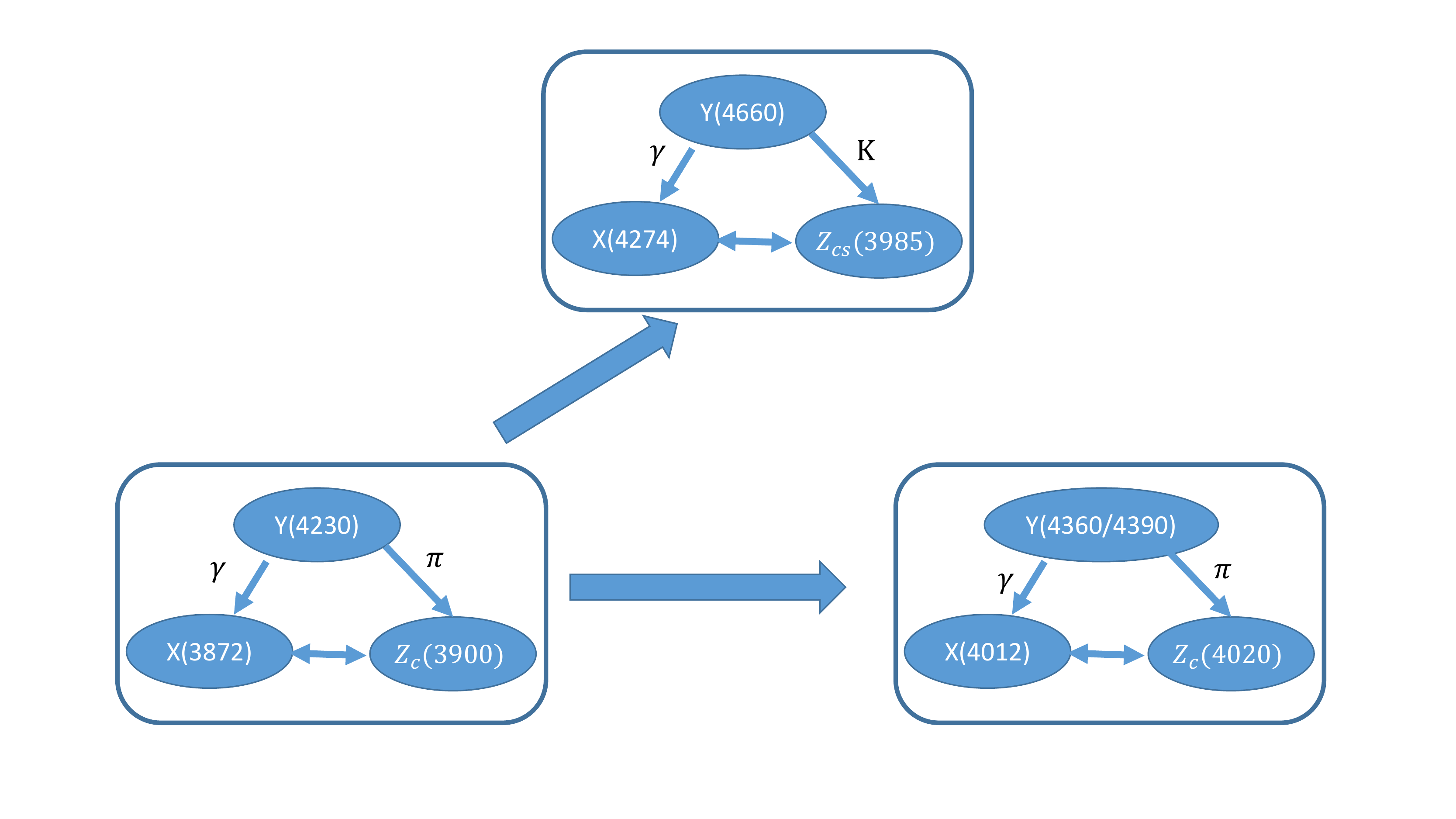}
\caption{Triple triangles for nine XYZ states.}
\label{fig:tt}
\end{figure}

\begin{figure}[htb]
\includegraphics[width=0.95\textwidth]{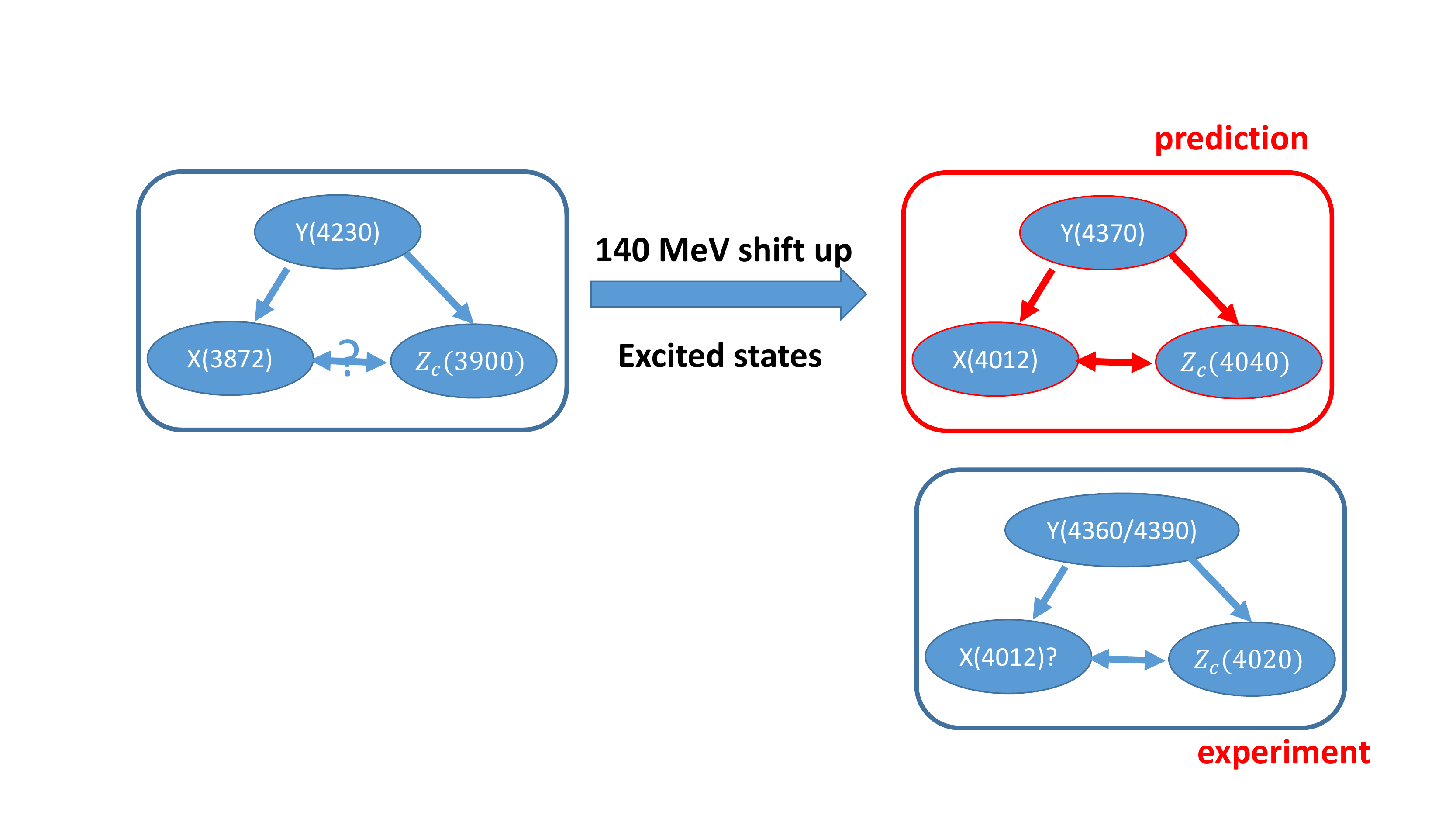}
\caption{The prediction and experimental observation of the first excited triangle of XYZ states.}
\label{fig:2t}
\end{figure}

\begin{figure}[htb]
\includegraphics[width=0.95\textwidth]{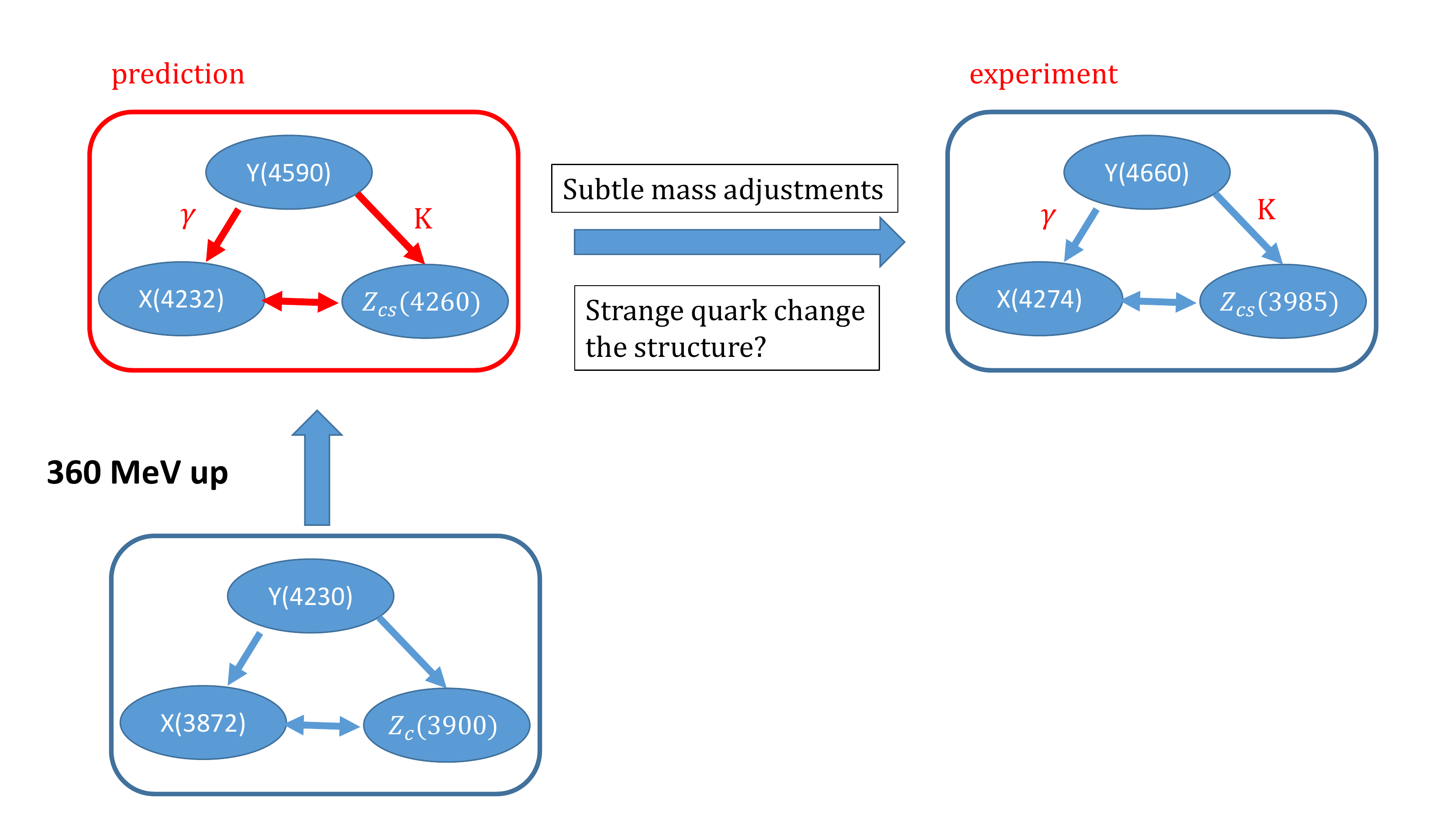}
\caption{The prediction and experimental observation of the second excited triangle of XYZ states involving with strange quark heavily.}
\label{fig:3t}
\end{figure}

Due to the open-charm thresholds and the masses of the well-known XYZ states $X(3872)$, $Y(4230)$, and $Z_c(3900)$~\cite{Zyla:2020zbs}, we assume that the three states construct the triple points of the fundamental triangle, and the radiative and hadronic transitions are illustrated as the triple sides as shown in the left-bottom of Fig.~\ref{fig:tt}. So starting from the fundamental triangle, i.e. the X(3872)-Y(4230)-Zc(3900) triangle, we can predict an excited triangle with the mass shift between $4.29$ $\mathrm{GeV}$ and $4.43$ $\mathrm{GeV}$, {\it i.e.} $140$ $\mathrm{MeV}$. As shown in Fig.~\ref{fig:2t}, it is obvious the Y and Z states belonging to the predicted triangle can be associated to the experimental observations, Y(4360/4390) and Zc(4020), respectively. Here we assume Y(4360) and Y(4390) are actually one vector state due to their similar masses. However, there is no such evidence of the X state close to X(4012) experimentally. Only it is well matched with another theoretical prediction~\cite{1204.2790}. Based on the similar argument, it is also another triangle can be obtained by shift the masses up to $360$ MeV, the difference between thresholds of $4.29$ $\mathrm{GeV}$ and $4.65$ $\mathrm{GeV}$, as shown in Fig.~\ref{fig:3t}. Here the predicted X and Y states are perfectly matched with experimental results $X(4274)$ and $Y(4660)$. While the mass of predicted Z state is obviously larger than the recently observed $Z_{cs}(3985)$, this discrepancy may be caused by the replacement of the strange quark with up/down quark, that will change the internal structure of the multiple quark state then decrease the mass of $Z$ state. Notice the subtle shift is about $300$ MeV that is just the difference between the masses of kaon meson and pion meson. Another possibility is the $Z_{cs}(3985)$ is not the Z state should be filled in this triangle, so another heavier Z state deserves searching for with different channels. Till now, there is no evidence of the $X(4274)$, discovered in $B \to J/\psi \phi K$ process~\cite{Aaij:2016iza}, is observed via $Y(4660) \to \gamma X(4274)$, $X(4274) \to \phi J/\psi$. So a search on this radiative transition channel is deserved. Additionally, because mass estimation of $X(4274)$ is based on a very rough way, other $X$ structures seen in $B \to J/\psi \phi K$ cannot be excluded such as $X(4140)$, and so $X(4274)$ is only the one of possible candidates.


The previous predictions on the two excited triangles have implicitly applied part of the interaction mechanism proposed in this paper. The complete form of the interaction hypothesis can be summarized as the following two rules.

\begin{itemize}
\item  Shifts between triangles: the radial position or orbital momentum between $\ket{h}$ and $\ket{c\bar{c}}$ for excited triangles keep similar to the fundamental triangle, but the $\ket{c\bar{c}}$ changes between the fundamental and excited ones according to the corresponding open-charm thresholds.
\item Transitions within a triangle: the $\ket{c\bar{c}}$ keeps same, only the radial position or orbital momentum between $\ket{h}$ and $\ket{c\bar{c}}$ should change.
\end{itemize}

Due to the second rule mentioned above and the observed channels, we can see there are more channels deserving search for. Some of them with higher priority are listed in Table.~\ref{tab:chan}. Here almost the observed or searched results are from PDG~\cite{Zyla:2020zbs}, while the others from recent published results~\cite{2004.13788, 2003.03705, 2001.01156, 1911.00885, 2012.02682, 2011.13850, 2010.14415}.

\begin{table}[ph]
\tbl{The observed/searched and proposed channels for XYZ states. Here $X \eta_c$ represents $\eta \pi^+ \pi^- \eta_c $, $3\pi \eta_c $, $\pi^+ \pi^- \eta_c$, and $\gamma \pi^0 \eta_c$; $\psi_2$ is $\psi_2(3823)$; $D_1$ is $D_1(2420)$.}
{\begin{tabular}{ c  c  c }
\toprule
 States & Observed/Searched & Proposed \\
 \colrule
 $X(3872)$ & $\rho J/\psi$, $\omega J/\psi$, $D^0 \bar{D}^{*0}$, $\pi^0 \chi_{c1}$, $\gamma J/\psi$, $\gamma \psi'$, $\gamma D\bar{D}$, $\pi^0 D^0\bar{D}^0$ & $\pi \pi \eta_c$, $\pi \pi \chi_{cJ}$ \\
 $Y(4230)$ & $\omega \chi_{c0}$, $\pi \pi J/\psi$, $\pi \pi h_c$, $\pi \pi \psi'$, $\pi^+ D^0 D^{*-}$, $\eta J/\psi$, $\eta' J/\psi$, $\pi^+ \pi^- \chi_{cJ}$, $X \eta_c$ & $\pi^0 \psi'$, $\gamma \chi_{cJ}$, $D^*\bar{D}^*$  \\
 $Z_c(3900)$ & $\pi J/\psi$, $\rho \eta_c$, $D\bar{D}^*$, $\pi \eta_c$ & $\pi h_c$, $\pi \psi'$, $\pi \pi \chi_{c0}$ \\
\colrule
$X(4012)$ &  & $\pi \pi \psi'$, $\pi \pi \psi_2$, $\pi \pi h_c$, $\pi \pi D \bar{D}$ \\
$Y(4360/4390)$ & $\pi \pi \psi'$, $\pi \pi \psi_2$, $D_1\bar{D}$, $\pi \pi h_c$, $\pi \pi \psi''$, $\eta J/\psi$ & $\pi D^* \bar{D}^*$ \\
$Z_c(4020)$ & $\pi h_c$, $D \bar{D}^*$ & $\pi \psi'$, $\pi \psi_2$, $D \bar{D}^*$, $\pi D \bar{D}$ \\
\colrule
$X(4274)$ & $\phi J/\psi$ & $D_s^{(*)} \bar{D}_s^{(*)} $, $\eta \chi_{cJ}$, $\eta' \chi_{cJ}$ \\
$Y(4660)$ & $K K J/\psi$, $K K h_c$, $K K \psi'$, $K D^{(*)} D_s^{(*)}$ & $\phi \chi_{cJ}$, $\eta J/\psi$, $\eta' J/\psi$, $\eta \psi'$, $\eta' \psi'$ \\
$Z_{cs}(3985)$ & $D D_s^*$, $D^* D_s$ & $K J/\psi$, $K \eta_c$, $K h_c$, $K \psi'$, $K \chi_{c0}$ \\
\botrule
\end{tabular}
\label{tab:chan}}
\end{table}

Furthermore, based on Eqs.~\ref{eq:hxy},~\ref{eq:hz},~\ref{eq:cc} and the following partial width formula
\begin{equation}
\Gamma(X/Y/Z \to f_h f_c) \propto \left| \braket{f_h | h } \right|^2 \left| \braket{f_c | c \bar{c}}\right|^2 \;,
\end{equation}
where $f_h$ and $f_c$ indicate light hadrons and charmonia in the final states, we derive the formulae for relations of the partial widthes of XYZ states decaying into light hadrons plus hidden-charm final states:
\begin{equation}
\label{eq:xandy}
\frac{\Gamma(X \to f_{h1} f_{c1})}{\Gamma(X \to f_{h2} f_{c1})} = \frac{\Gamma(Y \to f_{h1} f_{c2})}{\Gamma(Y \to f_{h2} f_{c2})}
\end{equation}
and
\begin{equation}
\label{eq:zandy}
\frac{\Gamma(Z \to f_{h1} f_{c1})}{\Gamma(Z \to f_{h1} f_{c2})} = \frac{\Gamma(Y \to f_{h2} f_{c1})}{\Gamma(Y \to f_{h2} f_{c2})} \;.
\end{equation}
Here the phase space effect has been ignored since the masses of XYZ states in one triangle are similar to each other. For example, from Eq.~\ref{eq:xandy} and the assumption that $X(3872)$ and $Y(4230)$ are assigned in the same triangle, we can have
$$
\frac{\Gamma(X(3872) \to \pi \pi J/\psi)}{\Gamma(X(3872) \to \omega J/\psi)} = \frac{\Gamma(Y(4230) \to \pi \pi \chi_{c0})}{\Gamma(Y(4230) \to \omega \chi_{c0})} \;,
$$
where corresponding final light hadron or hidden-charm final states $\pi \pi$, $J/\psi$, $\omega$, and $\chi_{c0}$ have replaced $f_{h1}$, $f_{c1}$, $f_{h2}$, and $f_{c2}$, respectively. However, because available experimental measurements are very limited, only a few predictions are obtained. And we do not use the results from~\cite{2004.13788} since it only has a significance of $4.2$ standard deviation, and its results are consistent with the charged mode~\cite{1611.01317}, which provides more precise measurements of $e^+ e^- \to \pi \pi J/\psi$. And we do not use the results from~\cite{2012.02682,2011.13850, 2010.14415}, since unfortunately only upper limits of the Born cross sections are provided in these paper without the corresponding coupling strength that can be applied in our calculation directly. These predictions are listed in Table~\ref{tab:pred}.
\begin{table}[ph]
\tbl{The predictions of production or decay rates for XYZ states.}
{\begin{tabular}{ l  l  l }
\toprule
 Items & Prediction & Comment \\
\colrule
$B(Y(4230)\to \pi^+ \pi^- \chi_{c0})\Gamma_{ee}^{Y(4230)}$(in eV) & $2.3 \pm 0.9$ & with $Y(4230)\to \omega \chi_{c0}$ \\
\colrule
$B(Y(4230)\to \pi^+ \pi^- \chi_{c0})\Gamma_{ee}^{Y(4230)}$(in eV) & $<11.4$ & with $Y(4230)\to \gamma \chi_{c0}$ \\
\colrule
$\Gamma(Z_c(3900)\to \pi h_c)/\Gamma(Z_c(3900)\to \pi J/\psi)$ & $0.4 \sim 3.0$ & not seen in experiment \\
\colrule
$\Gamma(Z_c(3900)\to \pi \psi')/\Gamma(Z_c(3900)\to \pi J/\psi)$ & $0.1 \sim 1.0$ & not seen in experiment \\
\botrule
\end{tabular}
\label{tab:pred}}
\end{table}
All the obtained predictions are consistent with each other when different experimental inputs are used in the calculation. We notice both the values of $\Gamma(Z_c(3900)\to \pi h_c)/\Gamma(Z_c(3900)\to \pi J/\psi)$ and $\Gamma(Z_c(3900)\to \pi \psi')/\Gamma(Z_c(3900)\to \pi J/\psi)$ take large uncertainty. The reason is the experimental result of $Y(4230)\to \pi \pi J/\psi$ is used, while in this measurement multiple solutions are found~\cite{1611.01317}. Considering no evidence has been observed in experiments till now, the destructive solution is more favored than the constructive one for the multiple solutions. During our calculations we have also realized the negative results are worthy to reported, since the upper limits sometimes will transform into bound limits or other constraints due to the relations.

In summary, we have proposed triple triangle relations based on the tetra-quark component hypothesis $\ket{h}\ket{c\bar{c}}$, with the S-wave open-charm thresholds as the weights, to classify the exotic XYZ states. Except $X(4012)$, that has not been observed experimentally yet, all the other predicted states can mostly consistent with presently observed states. Notice this triangle relation proposed by us is not the well-known triangle singularity theory, and it is just experimental results driven. Within this frame, we also propose some channels deserving search and predict a few production/decay rates of them.  They are main results of this paper and listed in Tables~\ref{tab:chan} and ~\ref{tab:pred}. Hope further experimental discoveries and measurements will benefit from our results.

It should also be mentioned that the original $Z_{cs}$ states predicted by us is obviously heavier than recently discovered $Z_{cs}(3985)$. Further search in other channels such as $K J/\psi$ are deserved, even this mass difference may be explained by the mass difference between kaon and pion. We have not discussed the widthes of the XYZ states, as well as the decay rates of open-charm channels since they are should be significantly affected by the strong coupling to open-charm channel near thresholds. Till now, we don't have a convinced method to deal with it. But whose widthes should be similar to each other when the XYZ states are in one triangle due to their similar internal structure, except one state is below the open-charm threshold then its width would be obviously narrower. It means the width of $X(4012)$ would be dozens of MeV or narrower since it seems just below the $D^*D^*$ threshold. The number of S-wave open-charm thresholds is larger than the number of the triangles proposed in this paper due to limited experimental results, that implies there maybe undiscovered XYZ states and more searches are deserved. And some thresholds are close to each other, that should make the identification and classification of the XYZ states difficult. We have not discussed other possible excited states, who will take different quantum numbers and larger radial positions or higher orbital momentum between $\ket{h}$ and $\ket{c\bar{c}}$ , since the experimental results on the XYZ states are still too limited. These unknown states may expand the triangle relations to polygon relations in principle.

\section*{Acknowledgments}

{The author thanks Yuping Guo, Aiqiang Guo, and Wolfgang K\"{u}hn for inspiring discussions. \\
This work is supported in part by National Key Research and Development Program of China under Contract No. 2020YFA0406301.}


\end{document}